\documentclass[twocolumn,showpacs,pra,aps]{revtex4}

\usepackage{dcolumn}
\usepackage{bm}

\begin{document}

\title{A truth about Brownian motion in gases and in general}

\author{Yuriy E. Kuzovlev}
\email{kuzovlev@kinetic.ac.donetsk.ua} \affiliation{Donetsk Institute
for Physics and Technology (DonPTI NASU), 83114 Donetsk, Ukraine}


\begin{abstract}
Real thermal motion of gas molecules, free electrons, etc., at long
time intervals (much greater than mean free-flight time) possesses,
contrary to its popular mathematical models, essentially non-Gaussian
statistics. A simple proof of this statement is suggested basing on
only the determinism and reversibility of microscopic dynamics and
besides incidentally derived virial expansion of a path probability
distribution of molecular Brownian particle.
\end{abstract}


\maketitle

{\bf 1}. Thermal chaotic Brownian motion of particles of the matter
is mechanism of diffusion as well as many other transport processes.
It is interesting by itself too, since statistics of random walk of
test or marked ``Brownian'' particles contains complete information
about transport process as a whole including its noise and
fluctuations. In spite of this, dynamical theory of molecular
Brownian motion remains almost undeveloped. Apparently, the common
opinion is that anyway it should confirm the well known beautiful
probability-theoretic scheme based on the celebrated ``law of large
numbers'' \cite{jb} which foretells that at sufficiently large
spatio-temporal scales probability density of displacement, or path,
$\,\Delta \bm{R}\,$, of a Brownian particle (BP) during time $\,t\,$
has universal form of the Gaussian distribution. For symmetric 3-D
random walk that is

$\,V_G(t,\Delta\bm{R})\,=\,(4\pi
Dt)^{-3/2}\,\exp{(-\Delta \bm{R}^2/4Dt)}\,$\,\,.

In fact, it was rigorously proved \cite{sin,gal} that such is the
asymptotic of chaotic walk of hard ball in so non-random environment
as static periodic lattice of scatterers (also hard balls).
Seemingly, Brownian motion of gas atoms or molecules all the more can
not display something else. Indeed, the Boltzmann-Lorentz equation
\cite{re}, which comes from the Boltzmann equation when applied to
self-diffusion of gas atoms, produces the Gaussian asymptotic.
Nevertheless, not all is so simple as seems.

Recall, firstly, that the Boltzmann-Lorentz equation along with the
Boltzmann's one is rather rough model since results from violent
truncation of the exact BBGKY hierarchy of equations and its closure
with the help of archaic Boltzmann's ``molecular chaos'' hypothesis
\cite{re,bog,sil}. However, it was pointed out more than once
\cite{re,kac} that Boltzmann's argumentation \cite{bol} looks enough
convincing only if applied to spatially uniform gas. As to attempts
to deduce ``molecular chaos'' from BBGKY equations themselves, they
proved to be unsuccessful \cite{ol}. In opposite, it was shown
\cite{i1} (see also \cite{i2,p1}) that reformulation of BBGKY
equations in terms of inter-particle ``collisions'' (instead of
continuous interactions) leads to equations which are evidently
incompatible with this hypothesis if a gas is spatially nonuniform.
In other words, the very fact of participation of particles in the
same collision (or connected conglomerate of collisions) at indicated
place is sufficient reason for mutual statistical dependence of the
particles. At the same time, if wishing to consider displacement of
BP (test particle) one has to localize its initial position and
thereby disturb uniformity (translational invariance) of infinitely
many distribution functions (DF) what describe BP under its
interactions with gas. Consequently, after all one has to deal with
an infinite chain of equations. Attempts to construct approximate
solutions to them were made in works \cite{i1} (see also \cite{i2})
and \cite{p1}. Their findings, of course, qualitatively differ from
the Gaussian asymptotic (but qualitatively confirm the early
phenomenological theory \cite{pjtf,bk12,bk3}).

Secondly, Gaussian asymptotic of the ball's walk in static lattice is
closely connected with its ergodicity \cite{sin,gal,ar}: results of
time averaging over $\,n\hm\rightarrow\infty $ fragments of its
trajectory (with each fragment consisting of free flight and
collision) almost surely are independent on the trajectory. This is
possible due to the fact that any trajectory is fully determined by
its single fragment (e.g. initial one), and a ratio of number,
$\,d=5n\,$, of quantities describing the trajectory to number,
$p=5\,$, of its specifying parameters is unrestrictedly large,
$d/p\hm\rightarrow\infty \,$. But a (weakly non-ideal) gas is quite
another matter. Here any trajectory of Brownian (test) particle is
made of $\,n\approx t/\tau\,$ similar fragments, with $\,t\,$ being
observation time and $\,\tau\,$ mean free flight time, and is
exhaustively characterized by $\,d=6n\,$ quantities. However, it is
specified by not only initial state of the particle under
consideration but also initial states of many other particles. It is
easy to see \cite{p1} that a number $\,m\,$ of such particles grows
with time by far faster than $\,n\,$, so that $\,n/m\hm\rightarrow
0\,$. Thus now the ratio of number of quantities what completely
describe all details of a trajectory to number, $\,p=6m\,$, of
parameters (initial conditions) what determine these details, tends
to zero, $\,d/p\hm\rightarrow 0\,$, as contrasted with above case. By
this reason, hardly results of time averaging (for instance,
``diffusivity'' or ``collision probabilities'') will turn out to have
``almost sure'' same limits for different trajectories. Here are no
grounds for ergodicity, the ``law of large numbers'' (which needs in
beforehand prescribed  ``collision probabilities'' \cite{jb,i2,p3})
and Gaussian asymptotic. Even in finite gas of $\,N<\infty\,$
particles (in box or torus) necessary grounds for ergodicity arise
not earlier than after time $\,>N\tau\,$ \cite{p1}. This gives
illustration of footnote remark in \cite{ar} that a value of
ergodicity in physics looks overestimated since limit
$\,N\hm\rightarrow \infty $ is much more important for physics than
limit $\,t\rightarrow\infty \,$. Besides, in essence, such warning
was highlighted already by N.\,Krylov \cite{kr}.

In principle, similar reasonings \cite{i2,bk3,i3,i4}, as well as
BBGKY equations or their analogues, are applicable to any realistic
many-particle system, and therefore particular case of gas has
general physical significance. But BBGKY equations yet are so bad
investigated that none supplementary tools would be superfluous.
Below, basing on determinism and reversibility of microscopic
dynamics only, we will derive a kind of virial expansion for true
probability distributuion of the BP's displacement,
$\,V_0(t,\Delta\bm{R})\,$. Then connect coefficients of this
expansion with join correlation functions of BP and gas and with
usual join DF. Finally, show that all these taken together imply a
differential inequality for $\,V_0(t,\Delta\bm{R})\,$ which
definitely forbids Gaussian asymptotic of $\,V_0(t,\Delta\bm{R})\,$
(but allows for non-Gaussian distribution obtained in \cite{p1} by
approximative solving BGKY equations).

{\bf 2}.  Let a gas of $N\gg 1$ atoms in volume $\,\Omega $ contains
a ``Brownian particle'' (BP). Consider, under the thermodynamical
limit $N\rightarrow\infty $, $\Omega\rightarrow\infty $, $N/\Omega
\hm =\nu_{\,0}\hm = \,$\,\,const\,, statistical ensemble of phase
trajectories of this system which responds to canonical equilibrium
distribution of its initial state at $t=0$, supposing that at $t>0$
BP is subjected to constant external force $\bm{f}\,$. Time
reversibility of phase trajectories implies many relations between
statistical characteristics of their ensemble \cite{jetp1,jetp2,p}.
In particular, according to \cite{jetp1} one can write
\begin{equation}
\begin{array}{c}
\langle A(\bm{q}(t))B(\bm{q}(0))\,e^{-\,\mathcal{E}(t)/T}\,\rangle
=\langle B(\bm{q}(t))A(\bm{q}(0))\rangle\,\label{sim0}
\end{array}
\end{equation}
Here brackets $\langle ...\rangle\,$ designate ensemble averaging,
$\,\bm{q}(0)=\bm{q}=\{\bm{R},\bm{r}_1,...\,,\bm{r}_N\}$ are space
coordinates of BP and atoms, $\mathcal{E}(t)=\bm{f}\cdot
[\bm{R}(t)-\bm{R}(0)]$ is work made by the external force during time
interval $t\,$, $A(\bm{q})$ and $B(\bm{q})$ are ``arbitrary
functions'', and $\,T$ is initial temperature of the system. This
equality holds also when BP has internal degrees of freedom.

Let $\,B(\bm{q})=\Omega\,\delta(\bm{R}-\bm{R}_0)\exp{[-\sum_j
U(\bm{r}_j)/T\,]}$ and $A(\bm{q})=\delta(\bm{R}-\bm{R}^{\prime})\,$.
Then r.h.s. in (\ref{sim0}) takes form
\[
\begin{array}{c}
\langle B(\bm{q}(t))A(\bm{q})\rangle =\mathcal{F}\{t,
\bm{R}_0,\phi|\bm{R}^{\prime}\}\,\equiv\,
V_0(t,\bm{R}_0-\bm{R}^{\prime})\,+\\
+\sum_{n\,=1}^{\infty } (\nu_{\,0}^n/n!)\int^n F_n(t,\bm{R}_0,
\bm{r}_1\,...\,\bm{r}_n|\bm{R}^{\prime})\prod_{j\,=1}^n
\phi(\bm{r}_j)\,\,,
\end{array}
\]
where\, $\,\phi(\bm{r})=\exp{[-\,U(\bm{r})/T\,]}-1\,$,\, symbol
$\int^n $ means integration over $\bm{r}_1...\,\bm{r}_n\,$,
function\, $\,V_0(t,\bm{R}_0-\bm{R}^{\prime})\,$ is probability
density of finding BP at $t\geq 0\,$ at point $\bm{R}_0$, and
$\,F_n(t,\bm{R}_0,\bm{r}_1\,...\,\bm{r}_n|\bm{R}^{\prime})\,$
probability density of this event and simultaneously finding some
atoms at points $\bm{r}_j\,$, under condition that BP had started
from point $\bm{R}^{\prime}\,$. Clearly, \,\,
$\,V_0(0,\bm{r})=\delta(\bm{r})\,$, $\,\int V_0(t,\bm{r})\,d\bm{r}
=1\,$, and
\[
\begin{array}{c}
F_n(0,\bm{R}_0,\bm{r}_1...\,\bm{r}_n|\bm{R}^{\prime})=
\delta(\bm{R}_0-\bm{R}^{\prime})
\,F_n^{(eq)}(\bm{r}_1...\,\bm{r}_n|\bm{R}_0)\,\,, \\
\mathcal{F}\{0,\bm{R}_0,\phi|\bm{R}^{\prime}\}=
\delta(\bm{R}_0-\bm{R}^{\prime})\,
\mathcal{F}^{(eq)}\{\phi |\bm{R}_0\}\,\,,\\
\mathcal{F}^{(eq)}\{\phi |\bm{R}_0\}\,\equiv \,1\,+ \\
+\sum_{n\,=1}^{\infty }(\nu_{\,0}^n/n!)\int^n
F_n^{(eq)}(\bm{r}_1...\,\bm{r}_n|\bm{R}_0)
\prod_{j\,=1}^n \phi(\bm{r}_j)\,\,,
\end{array}
\]
where $F_n^{(eq)}(\bm{r}_1...\,\bm{r}_n|\bm{R}_0)$ are equilibrium DF
and $\mathcal{F}^{(eq)}\{\phi |\bm{R}_0\}$ their generating
functional. In respect to atoms all the DF are normalized as usually
\cite{bog,sil}, that is $F_n(...\,\bm{r}_k...\,|\bm{R}^{\prime})
\rightarrow F_{n-1}(...\bm{r}_{k-1},
\bm{r}_{k+1}...\,|\bm{R}^{\prime})\,$ if $\,\bm{r}_k\rightarrow
\infty\,$, and \,$F_1^{(eq)}(\bm{r}_1|\bm{R}^{\prime}) \rightarrow
1\,$ if $\,\bm{r}_1\rightarrow \infty\,$. At one this requirement
expresses the ``principle of decay of correlations''.

Initial DF
$\,F_n(0,\bm{R}_0,\bm{r}_1\,...\,\bm{r}_n|\bm{R}^{\prime})\,$ include
all equilibrium correlations between BP and atoms. But any evolution
of (ensemble of states of) the system is conjugated with disturbance
of detailed balance of collisions and therefore births additional
correlations due to joint participation of particles in excess
collisions (or ``joint nonparticipation'' in missing ones). These
correlations could be qualified as non-equilibrium ones, but with
those reservation that sometimes they describe evolution of not so
much a system on its own as our information about it. Traditionally,
their contributions to DF are termed ``correlation functions'' (CF)
\cite{re,bog,sil,bal}. Let us designate them as
$\,V_n(t,\bm{R}_0,\bm{r}_1\,...\,\bm{r}_n|\bm{R}^{\prime})\,$ and
pick out them from $\,F_n(t>0,...)\,$ by definition as follows:
\[
\begin{array}{c}
\mathcal{F}\{t,\bm{R}_0,\phi|\bm{R}^{\prime}\}=
\mathcal{F}^{(eq)}\{\phi |\bm{R}_0\}\,
[\,V_0(t,\bm{R}_0-\bm{R}^{\prime})\,+\\ +\sum_{n\,=1}^{\infty }
(\nu_{\,0}^n/n!)\int^n V_n(t,\bm{R}_0,\bm{r}_1...\,\bm{r}_n|
\bm{R}^{\prime})\prod_{j\,=1}^n \phi(\bm{r}_j)]\,
\end{array}
\]
From here it is obvious that $\,V_n(0\,,...\,)=0\,$, and
$\,V_n(\,t\,,...)\rightarrow 0\,$ if at least one of the points
$\,\bm{r}_k\rightarrow \infty\,$. In particular,
\[
\begin{array}{c}
F_1(t,\bm{R}_0,\bm{r}_1|\bm{R}^{\prime})\, = \\
=V_0(t,\bm{R}_0-\bm{R}^{\prime})F_1^{(eq)}(\bm{r}_1|
\bm{R}_0)+V_1(t,\bm{R}_0,\bm{r}_1|\bm{R}^{\prime})\,\,\,,
\end{array}
\]
where function $\,V_1(t,\bm{R}_0,\bm{r}_1|\bm{R}^{\prime})\,$
comes from the usual ``pair correlation function''
\cite{sil,bal} in the full two-particle phase space:
\begin{equation}
\begin{array}{c}
V_1(t,\bm{R}_0,\bm{r}_1|\bm{R}^{\prime})=\int \int
V_1(t,\bm{R}_0,\bm{r}_1,\bm{P}_0,\bm{p}_1|\bm{R}^{\prime})\,
d\bm{P}_0\,d\bm{p}_1 \label{p1}
\end{array}
\end{equation}
The latter can be connected with the full DF exactly like
$\,V_1(t,\bm{R}_0,\bm{r}_1|\bm{R}^{\prime})\,$ is connected with
coordinate ones:
\begin{equation}
\begin{array}{c}
F_1(t,\bm{R}_0,\bm{r}_1,\bm{P}_0,\bm{p}_1|\bm{R}^{\prime})\,=\,
V_0(t,\bm{R}_0-\bm{R}^{\prime},\bm{P}_0)\times \label{cf1}
\\ \times\, F_1^{(eq)}(\bm{r}_1,\bm{p}_1|\bm{R}_0)+
V_1(t,\bm{R}_0,\bm{r}_1,\bm{P}_0,\bm{p}_1|\bm{R}^{\prime})
\end{array}
\end{equation}

Next, consider left side of (\ref{sim0}).  Multiply and divide it by
$\langle B(\bm{q})\rangle$\, and use the fact that any expression
$\langle \Phi\,B(\bm{q})\rangle /\langle B(\bm{q})\rangle\,$, with
$\Phi$ being some functional of system's phase trajectory, in view of
determinism of translation along any concrete trajectory can be
interpreted as $\,\Phi$'s\, average over new ensemble of trajectories
induced by new, non-equilibrium, ensemble of initial coordinates and
momentums of the system, namely, defined by probability distribution
$\,\rho (\bm{q},\bm{p})= B(\bm{q})\, \rho_0(\bm{q},\bm{p})/\langle
B(\bm{q})\rangle\,$, where $\,\rho_0(\bm{q},\bm{p})\,$ denotes
original equilibrium distribution. Noticing also that $\,\langle
B(\bm{q})\rangle =\mathcal{F}^{(eq)}\{\phi |\bm{R}_0\}\,$, we come to
\[
\begin{array}{c}
\langle A(\bm{q}(t))B(\bm{q}(0))\,e^{-\,\mathcal{E}(t)/T}\,\rangle
=\\ =\,V\{t,\bm{R}^{\,\prime}|\phi ,\bm{R}_0 \}\,\,e^{-\,\bm{f}\cdot
[\,\bm{R}^{\,\prime}-\,\bm{R}_0]/T}\, \mathcal{F}^{(eq)}\{\phi
|\bm{R}_0 \}\,\,\,,
\end{array}
\]
where $\,V\{t,\bm{R}^{\prime}|\phi ,\bm{R}_0 \}\,$ is probability
density of finding BP at time $\,t\,$ at point $\,\bm{R}^{\prime}\,$
ander conditions that initially it was placed at point
$\,\bm{R}_0\,$ while the gas was in such perturbed
spatially nonuniform non-equilibrium state which
would be equilibrium in presence of the potential $\,U(\bm{r})\,$.
Mean concentration of atoms in such state equals to
\[
\begin{array}{c}
\nu\{\bm{r}|\phi ,\bm{R}_0 \}\hm = [1+\phi(\bm{r})]\,\delta
\ln\mathcal{F}^{(eq)}\{\phi |\bm{R}_0 \}/\delta\phi(\bm{r})
\end{array}
\]

Thus, in total, we arrive at exact relation
\begin{equation}
\begin{array}{c}
V\{t,\bm{R}^{\prime}|{\phi ,\bf R}_0 \}\,\,e^{-\,\bm{f}
\cdot[\,\bm{R}^{\,\prime}-\,\bm{R}_0]/T}
\,=\,V_0(t,\bm{R}_0-\bm{R}^{\,\prime})\,+\label{r}\\
+\sum_{n\,=1}^{\infty } (\nu_{\,0}^n/n!)\int^n
V_n(t,\bm{R}_0,\bm{r}_1...\,\bm{r}_n|\bm{R}^{\,\prime})
\prod_{j\,=1}^n \phi(\bm{r}_j)\,\,\,
\end{array}
\end{equation}
which connects probability distribution of BP's path in initially
non-equilibrium nonuniform gas and analogous distribution, together
with generating functional of ``non-equilibrium correlations''
between previous path of BP and its current environment, for
initially equilibrium uniform gas. In case $\,\bm{f}=0\,$, therefore,
r.h.s. represents wholly equilibrium Brownian motion.  Notice that
similar relation, for perturbed gas density in place of the BP's
distribution, was considered in \cite{p3}.

{\bf 3}. To understand benefits of this relation, it is useful to
consider it under the formal ``Boltzmann-Grad limit'' \cite{ol} when
gas density increases, $\,\nu_{0}\rightarrow\infty\,$, while radii of
(short-range repulsive) interactions between BP and atoms, $\,r_b\,$,
and inter-atomic, $\,r_a\,$, decrease in such way that ``gas
parameters'' $\,r_a^3\nu_0 \,$ and $\,r_b^3\nu_0 \,$ go to zero but
mean free paths of atoms, $\,\lambda\hm =(\pi r_a^2\nu_0 )^{-1}\,$,
and mean free path of BP, $\,\Lambda =(\pi r_b^2\nu_0 )^{-1}\,$, stay
fixed. Thus one obtains ``ideal weakly-non-ideal gas'' where
$\,F_n^{(eq)}\rightarrow 1\,$, in those sense that, for example,
$\,\nu_0\int [\,F_1^{(eq)}(\bm{r}_1|\bm{R}_0)-1]
 \,d\bm{r}_1\,\sim\,$ $\sim\, \nu_0 r_b^3\,\rightarrow 0\,$\,.
As the cosequence,
\[
\begin{array}{c}
\nu\{\bm{r}|\phi ,\bm{R}_0 \}\,\rightarrow \,
\nu_{\,0}\,[\,1+\phi(\bm{r})\,]\,=\,\nu_{\,0}\,
\exp{[-U(\bm{r})/T\,]}
\end{array}
\]

This simplification makes it quite obvious that in the Boltzmann-Grad
limit on left side of (\ref{r}) effects of the density perturbation
depend on its relative measure $\,\phi(\bm{r})\,$ only but not on
$\,\nu_0\,$. Hence, r.h.s. of (\ref{r}) also depends on
$\,\phi(\bm{r})\,$ only, that is at any given function
$\,\phi(\bm{r})\,$ (and fixed $\,\Lambda\,$ and $\,\lambda\,$) there
exist finite limits
\[
\begin{array}{c}
\lim\,\,\nu_{\,0}^n\int^n V_n(t,\bm{R}_0,\bm{r}_1...\,
\bm{r}_n|\bm{R}^{\prime})\prod_{j\,=1}^n \phi(\bm{r}_j)\,
\neq 0,\,\infty
\end{array}
\]

Moreover, choosing $\,\phi(\bm{r})=\phi =\,$const\, inside
sufficiently large sphere (for instance, $\,|\bm{r}-\bm{R}_0|\hm <v_s
t_0\,$ with $\,v_s\,$ being sound speed in our gas and $\,t_0>t\,$),
such that a fortiori none correlations go beyond it, we can conclude
that (at fixed $\,\Lambda\,$, $\,\lambda\,$ and $\,t\,$) there exist
limits
\begin{equation}
\begin{array}{c}
\lim\,\,\nu_{\,0}^n\int^n V_n(t,\bm{R}_0,\bm{r}_1...\,
\bm{r}_n|\bm{R}^{\prime})\,=\,V_n(t,\bm{R}_0-\bm{R}^{\prime})\,
\label{clim}
\end{array}
\end{equation}
At that, the gas described by left side of (\ref{r}) effectively,
from the point of view of BP, is spatially uniform at start and what
is more during any given time of keeping BP under observation. But
its density is $\,(1+\phi)\,$ times greater than density of the
right-hand gas. Correspondingly, (\ref{r}) takes the form of simple
virial expansion:

\begin{equation}
\begin{array}{c}
V_0(t,-\Delta\bm{R}\,;1+\phi )\,\,e^{\,\bm{f}\cdot\Delta \bm{R}/T}
\,=\,\,\,\,\,\,\,\,\,\,\,\,\,\,\,\,\,\,\,\,\,\,\,\,\\
\,\,\,\,\,\,\,\,\,\,\,\,=\,\sum_{n\,=\,0}^{\infty } \,\phi^n
\,V_n(t,\Delta\bm{R}\,;1)/n!\, \label{r2}
\end{array}
\end{equation}
Here we introduced $\,\Delta \bm{R}=\bm{R}_0-\bm{R}^{\prime}\,$ and
third argument representing (dimensionless) gas density measured in
units of $\,\nu_0\,$, so that $\,V_n(...\,;1)=V_n(...\,)\,$.

As it follows from (\ref{clim})-(\ref{r2}), in weakly nonideal gas
spatial non-equilibrium correlations do not disappear in relative
sense, i.e. if counting upon (elementary volume $\,\nu_0^{-1} \,$
what fall at) one atom, although they disappear in literal sense:
$\,V_n(t,\bm{R}_0,\bm{r}_1...\,\bm{r}_n| \bm{R}^{\prime})\rightarrow
0\,$.

Let us interpret this statement on the base of those few things
what are known about the correlation functions
\cite{bog,sil,i1,i2,p1,bal}, first of all,
$\,V_1(t,\bm{R}_0,\bm{r}_1,\bm{P}_0,\bm{p}_1|\bm{R}^{\prime})\,$.
In respect to $\,\bm{r}_1-\bm{R}_0\,$ this pair CF is nonzero
inside a ``collision cylinder''  having radius $\,r_b\,$ and
directed in parallel to relative velocity of the pair of particles
(BP and some atom) where absolute vlaue of the CF is
comparable with product of one-particle DF, i.e. with
first term on r.h.s. of (\ref{cf1}). However, the approximative
theory \cite{sil,bal}, conventionally accepted as satisfactory
one, says nothing about length of the cylinder.
On the other hand, formula (\ref{clim}) clearly prompts
that in rigorous theory effective length of collision cylinder
is finite and has size of order of  $\,\Lambda\,$.
Indeed, let it be $\,c_1\Lambda\,$,  then volume
of the cylinder equals to
\,$\,\pi r_b^2\,c_1\Lambda =\,c_1\nu_0^{-1} \,$, and
\begin{equation}
\begin{array}{c}
\int V_1(t,\bm{R}_0,\bm{r}_1,\bm{P}_0,\bm{p}_1|
\bm{R}^{\prime})\,d\bm{r}_1\, = \,\,\,\,\,\,\,\,\,\,\,\,\,\,\,\,\,\,\\
\,\,\,\,\,\,\,\,\,\,\,\,\,\,\,\,\,\,=
\,c_1\,\nu_0^{-1}\,\overline{V}_1(t,\Delta \bm{R},\bm{P}_0,
\bm{p}_1)\,\,\,,\label{vm}
\end{array}
\end{equation}
where\,
$\,\overline{V}_1(t,\Delta \bm{R},\bm{P}_0,\bm{p}_1)\,$\,
represents mean value of
$\,V_1(t,\bm{R}_0,\bm{r}_1,\bm{P}_0,\bm{p}_1|\bm{R}^{\prime})\,$
within the cylinder.  Combining this expression with
(\ref{p1}) and (\ref{clim}), at $\,n=1\,$, we make certain of
finitness of the limit:
\begin{equation}
\begin{array}{c}
V_1(t,\Delta \bm{R})\,=\,c_1\,\int \int \overline{V}_1(t,\Delta
\bm{R},\bm{P}_0,\bm{p}_1)\,d\bm{P}_0\, d\bm{p}_1 \label{mc}
\end{array}
\end{equation}
At the same time, in literal sense formulas (\ref{p1}) and (\ref{mc})
yield \,$\,V_1(t,\bm{R}_0,\bm{r}_1|\bm{R}^{\prime})\hm\sim \,
V_1(t,\Delta \bm{R})\,r_b^2/|\bm{r}_1-\bm{R}_0|^2\hm\sim \,
\nu_0^{-1}\,V_1(t,\Delta \bm{R})/(\Lambda|\bm{r}_1-\bm{R}_0|^2)\hm
\,\rightarrow \,0\,$\,\,,\, since under integration over momentums a
very small part of differently oriented collision cylinders only
covers (fixed) point  $\,\bm{r}_1\,$.

{\bf 4}. We have come to curious conclusions.

Operating with coordinate $\,\bm{r}_1\,$, let us average identity
(\ref{cf1}) over the collision cylinder:
\[
\begin{array}{c}
\overline{F}_1(t,\Delta \bm{R},\bm{P}_0,\bm{p}_1)\,=\,\\
=\,V_0(t,\Delta \bm{R},\bm{P}_0)\,G^{(eq)}(\bm{p}_1)\,+
\,\overline{V}_1(t,\Delta \bm{R},\bm{P}_0,\bm{p}_1)\,\,
\end{array}
\]
Here  $\,G^{(eq)}(\bm{p}_1)\,$  is the Maxwellian distribution,
and the result,
$\,\overline{F}_1(t,\Delta \bm{R},\bm{P}_0,\bm{p}_1)\,$,
can be treated as ensemble average of density of number
of two-particle configurations matching up collisions \cite{i1}.
Then, performing integration over momentums and
applying formula (\ref{mc}), we obtain
\begin{equation}
\begin{array}{c}
W_1(t,\Delta \bm{R})\,\equiv\, \int\int \overline{F}_1(t,\Delta
\bm{R},\bm{P}_0,\bm{p}_1)\,d\bm{P}_0\, d\bm{p}_1\,=\\
=\, V_0(t,\Delta \bm{R})\,+\,c_1^{-1}\,V_1(t,\Delta
\bm{R})\,\,\label{cf2}
\end{array}
\end{equation}

For simplicity, confine ourselves by the case $\,\bm{f}=0\,$. From
relation (\ref{r2}) whose both sides in this case concern equilibrium
(hence, spherically symmetric) Brownian motion, we can connect \,
$\,V_1(t,\Delta\bm{R})\hm =V_1(t,\Delta\bm{R}\,;1)\,$ with derivative
of the left side by the relative gas density: $\,V_1(t,\Delta
\bm{R})\,\hm =\, [\,\partial V_0(t,\Delta \bm{R}\,;1+\phi )/\partial
\phi\,]_{\,\phi\,= \,0}\,$\,.\, Then, combining this equality with
(\ref{cf2}) and noticing that functions $\,\overline{F}_1\,$ and
$\,W_1\,$ by their birth from DF $\,F_1\,$ are non-negative, in view
of  $\,W_1\geq 0\,$ we come to inequality
\begin{equation}
c_1 V_0(t,\Delta \bm{R}\,;1)+\left [\frac {\partial V_0(t,\Delta
\bm{R}\,;1+\phi )}{\partial \phi }\,\right ]_{\,\phi\,= \,0}\,\geq 0
\label{r3}
\end{equation}

Further, let us assume that at  $\,t\gg \tau =\Lambda /v_0\,$, where
$\,\tau \,$ is mean free flight time and  $\,v_0\,$ characteristic
thermal velocity of BP, a probability distribution of BP's path tends
to the Gaussian one (see paragraph 1): $\,V_0(t,\Delta
\bm{R})\rightarrow V_G(t,\Delta\bm{R})\,$\,. Besides, take into
acount that diffusivity of BP, $\,D\,$, in the Boltzmann-Grad gas is
arranged as $\,D=\Lambda v_0=v_0^2\tau \,$ and, consequently,
together with $\,\Lambda \,$ is inversely proportional to gas
density. Therefore left side of (\ref{r2}) should follow from
$\,V_G(t,\Delta\bm{R})\,$ after change $\,\Lambda \rightarrow \Lambda
/(1+\phi )\,$, $\,D \rightarrow D/(1+\phi )\,$, that is
\[
V_0(t,\Delta \bm{R}\,;1+\phi )\,\rightarrow \,\left (\frac
{1+\phi}{4\pi Dt}\right )^{3/2}\exp{\left [-(1+\phi)\,\frac {\Delta
\bm{R}^2}{4Dt}\right ]}\,\,
\]
Insertion of this expression to inequality (\ref{r3}) yields
\,$\,V_G(t,\Delta\bm{R})\,[c_1+3/2\hm -\Delta \bm{R}^2/4Dt\,] \,\geq
0\,$\,\,, i.e. produces obvious contradiction.

Hence, the theory does not allow for the Gaussian asymptotic! This
statement, or rather its new proof since it itself is not novell (see
paragraph 1), is principal result of the present paper.

Of course, it is not rigorous because we have managed without a
concrete definition of the averaging procedure marked by the
over-line (though heuristically this is advantage of our
consideration). But, in view of (\ref{vm}), this procedure may
influence upon numerical value of the coefficient $\,c_1\,$. This
notion, nevertheless, does not spoil our result, since it means
merely that we should substitute to (\ref{r3}) minimum of possible
values. Our presumption that $\,c_1\,$ is indeed a coefficient, but
not a function of $\,t,\Delta \bm{R},\bm{P}_0\,$ and $\,\bm{p}_1\,$
in (\ref{vm}) or $\,t\,$ and $\,\Delta \bm{R}\,$ in (\ref{mc}), also
can be justified. Physically, the length of collision cylinder,
$\,c_1\Lambda\,$, for pair of particles is bounded by collisions with
``third particles''  which knock out the pair from a number of
candidates for collisions. Since the ``third particles'' originate
mainly from uniform background, the constancy of $\,c_1\,$ seems
quite natural.

{\bf 5}. In conclusion, a few more remarks.

First, the above formulation of the Boltzmann-Grad limit implied that
size of BP is comparable with size of atoms. Therefore conclusions of
previous paragraph do not apply to ``macroscopic'' BP (in respect to
it, at least in the limit of infinitely large BP's mass, Gausian
asymptotic remains above suspicion).

Second, and what is allowed for by the inequality (\ref{r3})? Let
$\,V_0(t,\Delta \bm{R})\,$ at $\,t\gg \tau \,$ is characterized,
similarly to $\,V_G(t,\Delta \bm{R})\,$, by a single parameter, that
is diffusivity:\, $\,\,V_0(t,\Delta \bm{R}) \rightarrow
(2Dt)^{-3/2}\,\Psi(\Delta \bm{R}^2/2Dt)\,$\,\,, where $\,\int
\Psi(\bm{a}^2)\,d\bm{a}\,=1\,$. Then inequality (\ref{r3}) reduces to
\[
\begin{array}{c}
(c_1+3/2)\Psi(x) +x\,d\Psi(x)/dx \,\geq \,0\,\,,
\end{array}
\]
thus requiring that function $\,\Psi(x)\,$ must have power-law long
tail: \, $\,\Psi(x)\propto\, x^{-\,\alpha }\,$\, at
$\,x\rightarrow\infty\,$, where, clearly, $\,\alpha\leq c_1+3/2\,$\,.
But such behavior would lead to unboundedness of higher-order
statistical moments of BP's path. Recall, however, that we have once
more parameter, $\,v_0\,$, and hence, at $\,t\gg \tau\,$, the
possibility to write \, $\,\,V_0(t,\Delta \bm{R})\rightarrow
(2Dt)^{-3/2}\, \Psi(\Delta \bm{R}^2/2Dt)\,\Theta (|\Delta
\bm{R}|/v_0t)\,$\,, where $\,\Theta(0)=1\,$ and $\,\Theta(x)\,$
enough rapidly decreases at infinity. From the point of view of
(\ref{r3}), this is also allowed variant. And it reproduces, if take
$\,c_1=2\,$, the asymptotic of distribution of BP's path found in
\cite{p1} in the frame of ``collisional approximation'' to BBGKY
equations \cite{i1,i2}.  At that, the role of BP was played by test
(marked) atom. Such the asymptotic reflects non-ergodicity of its
Brownian trajectories, which can be characterized as ``flicker''
fluctuations of diffusivity (and mobility) of BP \cite{i1,bk3,dev}.

It should be underlined that our present formalism and one exploited
in \cite{p1} have nothing common among them (except their
application). Therefore the remarkable nearness of their results can
be rated as evidence for adequacy of both approaches.

At last, thirdly, in our above consideration by the example of pair
DF we in fact have outlined connections between the correlation
functions, on one hand, and the specific distribution functions (for
pair and many-particle collisional configurations) first introduced
in \cite{i1}, on the other hand. In particular, function
$\,\overline{F}_1\,$ here is equivalent of function $\,A_2\,$ from
\cite{i1}, while function $\,W_1\,$ defined in (\ref{cf2}) is
equivalent of  $\,W_2\,$ from \cite{p1}. Extension of these
connections to higher-order DF will unify the two formalisms and,
undoubtedly, open new applications of the virial expansions (\ref{r})
and (\ref{r2}).

I would like to acknowledge Dr. I.\,Krasnyuk for many useful
conversations.


\begin{thebibliography}{25}
\bibitem{dev}
The present paper is devoted to 25-th anniversary of first work by me
(in collaboration with G.\,N.\,Bochkov), \cite{pjtf}, on the subjects
under consideration.
\end{thebibliography}
\end{document}